\documentclass[a4paper,11pt]{article}
\usepackage{pos}
\usepackage{booktabs}
\usepackage{graphicx}
\usepackage{subcaption}
\usepackage{setspace}
\usepackage{lineno}
\usepackage{float} 
\usepackage{tabularx}
\usepackage{enumitem}
\usepackage{wrapfig}

\usepackage{siunitx}
\DeclareSIUnit{\year}{yr}
\DeclareSIUnit{\EeV}{EeV}

\title{Measurement of the cosmic-ray flux from 2.5 EeV by the Pierre Auger Observatory after 19 years of operation }
 \ShortTitle{Auger spectrum}

\author*[a]{Diego Ravignani}

\affiliation[a]{ITeDA (CNEA/CONICET/UNSAM), Argentina}

\onbehalf{for the Pierre Auger Collaboration$^c$}
\affiliation[c]{Observatorio Pierre Auger, Av.\ San Mart{\'\i}n Norte 304, 5613 Malarg\"ue, Argentina\\
Full author list: {\rm\url{https://www.auger.org/archive/authors_icrc_2025.html}}}

\emailAdd{spokespersons@auger.org}

\abstract{
We present the spectrum of cosmic rays with energies above \SI{2.5}{\EeV} measured at the Pierre Auger Observatory after 19 years of operation, covering the period before the AugerPrime upgrade. Two independent event sets from the surface array of 1500 m-spaced detectors are combined, yielding a total exposure of \SI{104900}{\square\kilo\meter\steradian\year}. The first set includes events with zenith angles less than \SI{60}{\degree}, while the second consists of events between \SI{60}{\degree} and \SI{80}{\degree}, for which azimuthal asymmetries must be accounted for in the energy estimator. The threshold energy is chosen to ensure a trigger efficiency of the surface detector greater than 97\%, thus minimizing composition biases. The energy scale is determined using high-quality fluorescence measurements, providing calorimetric estimates without reliance on simulations.

A statistically successful combination is achieved within the uncorrelated systematic uncertainties of the individual spectra. All spectra are consistent when analyzing potential declination dependences, except for a mild modulation expected from the previously reported dipolar anisotropy. In particular, this statement applies to the northernmost declination band [+\SI{24.8}{\degree},+\SI{44.8}{\degree}], where only events with zenith angles between \SI{60}{\degree} and \SI{80}{\degree} contribute. Beyond the firmly established ankle and suppression spectral features, the combined spectrum across declinations -\SI{90}{\degree} to +\SI{45}{\degree} provides a measurement of the instep feature with more than 5$\sigma$ confidence.
}

\ConferenceLogo{PoS_ICRC2025_logo}

\FullConference{%
39th International Cosmic Ray Conference (ICRC2025)\\
15 -- 24 July, 2025\\
Geneva, Switzerland}

\begin{document}
\maketitle

\section{Introduction}

The Pierre Auger Observatory~\cite{PierreAuger:2015eyc} has accumulated, since its beginning in 2004, the largest dataset of ultra-high-energy cosmic rays currently available.
The volume and quality of the acquired data allowed us to measure a dipolar modulation of $\sim 6\%$ of an otherwise isotropic distribution of the arrival directions of cosmic rays with a significance greater than 5$\sigma$~\cite{PierreAuger:2024fgl}. 
Motivated by the observed dipole, we previously searched for a modulation of the spectrum with declination~\cite{PierreAuger:2020qqz, PierreAuger:2020kuy}.
We recently conducted a more comprehensive search exploiting a $\sim66\%$ larger exposure with a dataset that also includes events arriving with a zenith angle between \SI{60}{\degree} and \SI{80}{\degree} thus extending the declination reach from +\SI{24.8}{\degree} to +\SI{44.8}{\degree}~\cite{PierreAuger:2025hnw} that is also reported here. 

We present an energy spectrum using events observed by the water-Cherenkov detectors of the Surface Detector Array spaced at \SI{1500}{\meter} during the 19 years of the Phase I data taken between January 1, 2004 and January 1, 2023.
After this successful operation, the Observatory was upgraded, within a project named AugerPrime, with the goal of improving the measurement of the mass composition of ultra-high energy cosmic rays~\cite{Schmidt:2025}.
The upgrade includes new scintillator and radio detectors added in each position of the Surface Detector Array. 
We also upgraded the detector electronics to improve the timing and signal resolution of water-Cherenkov detectors and extended their dynamic range by adding a new PMT, smaller than the previous ones. 
We installed underground muon detectors at the positions of the \SI{433}{\meter} and \SI{750}{\meter} Surface Detector Arrays. 
Phase II data-taking started on April 1, 2023.
We show a preliminary spectrum built with events acquired during the first two years of Phase II.

\section{Vertical and inclined spectra}

The depth of the water-Cherenkov detectors grants them the geometric acceptance needed to measure particles arriving at large zenith angles.
However, the structure of the shower changes significantly for more vertical and inclined events. 
While the more vertical showers are symmetric around their axis and have a significant contribution at ground level of photons and electrons, the more inclined showers are composed predominantly of muons, and their axial symmetry is broken due to the geomagnetic field and other effects.  
Due to these reasons, we reconstruct the events arriving with a zenith angle up to \SI{60}{\degree}, denominated here as \emph{vertical} events, with a different method than those arriving between \SI{60}{\degree} and \SI{80}{\degree}, the \emph{inclined} events. 

We reconstruct vertical events using a function that describes the fall-off of detector signals with the distance to the shower axis~\cite{PierreAuger:2020yab}.
The value of this function at a distance of \SI{1000}{\meter} ($S(1000)$) provides a measure of the shower size.
We estimated a shower size independent of the zenith angle ($S_{38}$), equivalent to the size the shower would have had if it had arrived at a zenith angle of \SI{38}{\degree}, via an attenuation function, $S_{38} = S(1000) / f_\text{att}$, calculated with the Constant Intensity Cut method~\cite{Hersil:1961zz}. 
The attenuation function $f_\text{att}$ depends on the zenith angle through a variable $x= \sin^2\theta - \sin^2 \SI{38}{\degree}$ and on $S_{38}$ through a variable $ y = \log_{10}(S_{38} / S_0)$, with $S_0 = \SI{40}{VEM}$, and

\begin{equation}
    f_\text{att}(x,y) = 1 + \sum_{i=1}^{3} \sum_{j=0}^{2} a_{ij} \, x^i y^j,
    \label{eq:fatt}
\end{equation}

\noindent with parameters $a_{ij}$ in the Table~ \ref{tab:attenuation_vertical}.

\begin{table}[t]
\renewcommand{\arraystretch}{1.0}
\centering
\small
\caption{Coefficients of the attenuation function used to correct the shower size of vertical events.}
\begin{tabular}{@{} c c c c @{}} \toprule
  & $1$ & $y$ & $y^2$ \\ \midrule
$x$ & $-0.936$ & $ -0.005$ & $\phantom{+}0.4\phantom{00}$  \\
$x^2$ & $-1.62\phantom{0}$ & $-0.51\phantom{0}$ & $-0.13\phantom{0}$ \\
$x^3$ & $\phantom{+}0.92\phantom{0}$ & $-0.54\phantom{0} $ & $-1.75\phantom{0} $ \\ \bottomrule
\end{tabular}
\label{tab:attenuation_vertical}
\end{table} 

The shower size $S_{38}$ is calibrated with the energy measured with the telescopes of the Fluorescence Detector ($E_\text{FD}$) using high-quality events observed in coincidence with the Surface Detector Array.
From calibration data, we establish the relationship between the cosmic ray energy and the shower size, $E = A S_{38}^B$, with $A= \SI{186 \pm 3}{\peta\electronvolt}$, $B=1.021\pm 0.004$, and a correlation coefficient between $A$ and $B$, $\rho = -0.98$. 

We reconstruct inclined events by fitting a function that describes the signal pattern~\cite{PierreAuger:2014jss, PierreAuger:2015xho}.  
This map is scaled with the ratio of the measured and simulated shower size at \SI{10}{\EeV} ($N_{19}$).
The parameter $N_{19}$ attenuates with the zenith angle in a slightly model-dependent way, in particular, due to the residual contribution of photons and electrons.
This attenuation is corrected using the Constant Intensity Cut method as in the case of vertical events to derive an equivalent shower size at \SI{68}{\degree}, $N_{68} = N_{19} / f_\text{att}$.
The attenuation function is described as a simplified model of Eq.~\ref{eq:fatt}, with $x = \sin^2\theta - \sin^2 \SI{68}{\degree}$, $y=\log_{10} (N_{19})$, and $f_{\text{att}} = 1 + (0.292 - 0.468 \, y) \, x +  (-4.96 + 0.79 \, y) \, x^2$.
The shower size $N_{68}$ is calibrated in energy using events observed in coincidence with the Fluorescence Detector as with vertical events. 
The energy of inclined events is calculated from $E=A\, N_{68}^B$, with $A=\SI{5.29\pm 0.06}{EeV}$, $B=1.046\pm 0.014$, and a correlation coefficient $\rho_{AB} = -0.66$. 

We calibrated the events using a Fluorescence Detector reconstruction updated with an improved estimation of aerosol attenuation and parametrization of the longitudinal profile of light emission~\cite{Dawson:2019zva}. 
We calibrated the vertical and inclined events with the same fluorescence reconstruction so that both datasets share the same energy scale. 
The systematic uncertainty of the energy of vertical and inclined events is driven by the absolute energy scale of fluorescence detection (14\%)~\cite{Verzi:2013ajy} that is correlated between both data sets.
In addition, the uncertainty of the calibration parameters propagates as another source of systematic uncertainty in the energy but is not correlated between vertical and inclined events. 
Although the number of events available in the vertical calibration renders this source of systematic uncertainty negligible, inclined energies with fewer calibration events have a systematic uncertainty of about $2\%$.

For the spectrum based on vertical events, we selected events that have the detector with the highest signal surrounded by six working detectors to ensure the energy is well reconstructed. 
We applied a similar condition for inclined events, requiring the reconstructed core to be contained within a hexagon of six working detectors. 
We choose the energy thresholds of the vertical spectrum at $\log_{10}(E/\text{eV})=18.4$ ($E\sim\SI{2.5}{\EeV}$), and of the inclined spectrum at $\log_{10}(E/\text{eV})=18.6$ ($E\sim\SI{4}{\EeV}$), limits at which the trigger efficiency exceeds $97\%$~\cite{PierreAuger:2010zof}. 
Above the energy threshold, the exposure is constant and is calculated by monitoring the status of the detectors every~\SI{1}{\second}. 
From this thorough monitoring, we identified periods of unstable acquisition that accounted for less than $3\%$ of the observation time.
We excluded the events observed during these periods and penalized the exposure accordingly.  
We computed an exposure of \SI{81100 \pm 700}{\square\kilo\meter\steradian\year} for the vertical spectrum and \SI{23800 \pm 400}{\square\kilo\meter\steradian\year} for the inclined spectrum. 

\noindent
\begin{minipage}[t]{0.48\textwidth}
  \centering
  \includegraphics[width=\linewidth]{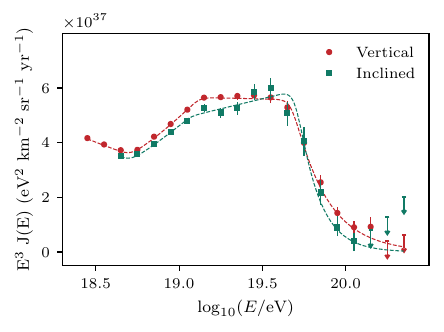}
  \captionof{figure}{Vertical spectrum using events arriving with a zenith angle less than \SI{60}{\degree}, and inclined spectrum from events with a zenith angle between \SI{60}{\degree} and \SI{80}{\degree}. The spectrum data corrected for the detector response and the fitted flux are shown. No cut in the declination of the arrival direction has been applied.}
  \label{fig:individual_spectra}
\end{minipage}
\hfill
\begin{minipage}[t]{0.48\textwidth}
  \centering
  \includegraphics[width=\linewidth]{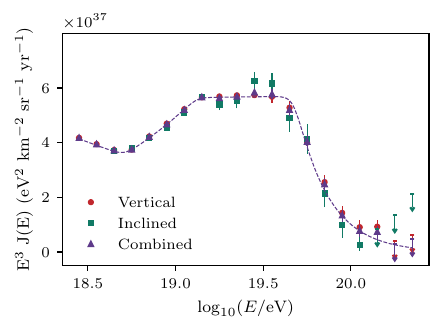}
  \captionof{figure}{Vertical and inclined spectra in their common declination band [-\SI{84.8}{\degree}, +\SI{24.8}{\degree}]. The inclined spectrum is shifted according to the recalibrated event energies calculated in the spectra combination. The resulting combined spectrum is shown.}
  \label{fig:combination}
\end{minipage}
\vspace{2em}

We fitted the spectrum data with a flux of cosmic rays modeled as a power-law function with smooth transitions,

\begin{equation}
\label{eq:flux}
J(E) = J_0 \left(\frac{E}{E_0}\right)^{-\gamma_0} \frac{\prod_{i=1}^3\left[1+\left(\frac{E}{E_{i}}\right)^{\omega_{i}^{-1}}\right]^{(\gamma_i-\gamma_{i+1})\,\omega_{i}}}{\prod_{i=1}^3\left[1+\left(\frac{E_0}{E_{i}}\right)^{\omega_{i}^{-1}}\right]^{(\gamma_i-\gamma_{i+1})\, \omega_{i}}},
\end{equation}

\noindent with the reference energy $E_0$ set at $10^{0.5} \sim \SI{3.16}{\EeV}$ and the three transitions widths fixed at $w_{i}=0.05$. We fitted the flux normalization $J_0$, the ankle energy $E_1$, the instep energy $E_2$, the fall-off energy $E_3$, and the four spectral indexes $\gamma_i$. The spectral index $\gamma_1$ corresponds to the region before the ankle, $\gamma_2$ between the ankle and the instep, and likewise for the other spectral indexes.

The observed number of events in each energy bin is affected, besides the prediction arising from the cosmic ray flux, by the effect of the detector response that includes the trigger efficiency, and the bias and finite resolution of the reconstructed energy~\cite{PierreAuger:2020qqz}.
We incorporated these effects in the spectrum fitting. 
We correspondingly corrected the observed flux in each bin by factors $c_i \sim [0.9 - 1]$, $J'_i = c_i \, J_i$, with $J_i = N_i / (\varepsilon \, \Delta E_i)$, with $N_i$ the observed number of events, and $\Delta E_i$ the energy difference between the bin limits.  
The vertical and inclined spectra corrected by the detector response and the corresponding fitted fluxes are shown in Fig.~\ref{fig:individual_spectra}.

\section{Dependency of the flux with declination}
\label{sec:declination}

The latitude of the Observatory at \SI{35.2}{\degree}~S permits the observation of cosmic rays arriving with a declination from the south celestial pole up to +$\SI{24.8}{\degree}$ with vertical events, and the range $[-\SI{84.8}{\degree}, +\SI{44.8}{\degree}]$ with inclined events.
We exploited the cumulative exposure of the vertical and inclined spectra by combining them into a single spectrum.
To guarantee the observation of the same sky, we combined them in their common field of view given by the declination band $[-\SI{84.8}{\degree}, +\SI{24.8}{\degree}]$ using the method in reference~\cite{PierreAuger:2021hun}.
We fitted the vertical and inclined data simultaneously with the flux model~(\ref{eq:flux}).  

\begin{figure}
    \centering
    \includegraphics{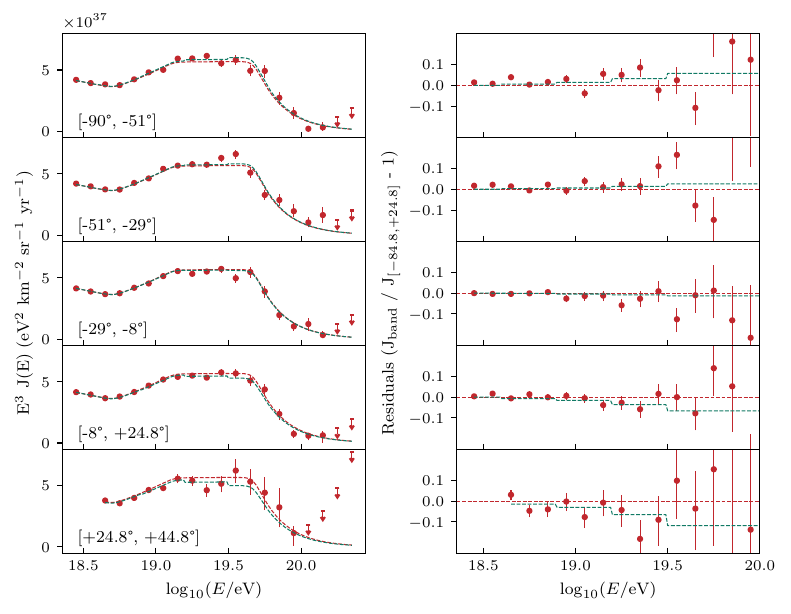}
    \caption{Left panel: Spectra in five declination bands combining vertical and inclined events. The flux fitted in the common declination band $[-\SI{84.8}{\degree}, +\SI{24.8}{\degree}]$ is shown in red, and the flux modulated by the dipole of the arrival direction distribution is shown in green.
    Right panel: Residuals of the combined spectra and the fitted flux in each declination band clipped to $[-0.25, 0.25]$. \label{fig:declination_bands}}
\end{figure}

During the fit, we included the effect of the uncorrelated systematics on the energy of the vertical and inclined events.
We fixed the energies resulting from the calibration for the vertical events as they have negligible systematics.
However, we fluctuated the energy of inclined events by varying the parameters $A$ and $B$ around the calibration estimates.
We use different $B$ deviations for energies below and above \SI{10}{\EeV}, and denominated them $\delta B$ and $\delta C$, respectively, to account for possible deviations of the inclined calibration from a pure power-law model originated, for example, by the evolution of the primary composition.
The combination maximizes a Poisson likelihood that contains contributions predicting the expected number of events in each bin of the vertical and inclined spectra.
In addition, during the fit, we penalized the deviations from the calibration estimates ($\delta A$, $\delta B$, and $\delta C$). 

We fitted simultaneously the eight spectral parameters and the three calibration parameters. 
The fitted deviations are $\delta A = \SI{160 \pm 39}{\peta\electronvolt}$ , $\delta B = 0.003 \pm 0.016$, and $\delta C = -0.02 \pm 0.02$.
The recalibration of inclined events increases the energy by $2.9\%$ at \SI{4}{\EeV}, $3.1\%$ at \SI{10}{\EeV}, and decreases it $1\%$ at \SI{100}{\EeV}. 
For the combination we obtained a deviance $D=40.5$, for which we estimated a $p\text{-value} \simeq 0.12$.
 We show in Fig.~\ref{fig:combination} the fitted flux and the combined spectrum calculated in each bin as, 

\begin{equation}
\label{eq:combined_spectrum}
J_i = \frac{c_i \, n_i + c_i' \, n_i'}{(\varepsilon + \varepsilon') \, \Delta E_i}, 
\end{equation}

\noindent with the unprimed and primed symbols corresponding to the vertical and inclined spectra. 
    
We searched for a dependence of the spectrum with declination by dividing the sky into five declination bands. 
We divided the range observed with vertical events $[-\SI{90}{\degree}, +\SI{24.8}{\degree}]$ in four bands of similar exposure. 
In addition, we considered a northern band $[+\SI{24.8}{\degree}, +\SI{44.8}{\degree}$] observed only with inclined events.
We compared the spectra in each declination band with the flux in the common declination range $[-\SI{84.8}{\degree}, +\SI{24.8}{\degree}]$.
We calculated the combined spectrum in each band as per Eq.~\ref{eq:combined_spectrum} using the number of events and exposures of each band.
We show the spectra in declination bands and the flux in the left panel of Fig.~\ref{fig:declination_bands}. 
We also show the flux expected in each band when the modulations given by the dipole in the arrival directions observed by Auger are considered. 
We show in the right panel of Fig.~\ref{fig:declination_bands} the residuals of the spectra in declination bands with respect to the fitted flux and the deviations of the flux modulated by the dipole.
The residuals follow the trend imprinted by the dipole between \SI{4}{\EeV} and \SI{32}{\EeV}, and statistical uncertainties dominate beyond \SI{32}{\EeV}.
The agreement of the dipole modulation with the spectra in each band is expected since the anisotropy and spectrum data sets overlap considerably.   

\begin{table}
\renewcommand{\arraystretch}{1.0}
\centering
\small
\caption{Spectral parameters of Phase I spectrum including the energy of the ankle ($E_1$), the instep ($E_2$), and the flux suppression ($E_3$), and the spectral indexes $\gamma$.}
\begin{tabular}{@{} l r r r p{5mm} l r r r @{}} \toprule
\multicolumn{1}{c}{\textbf{Parameter}} & \multicolumn{1}{c}{\textbf{Value}}  & \multicolumn{1}{c}{\textbf{\boldmath$\sigma_{\text{stat}}$}}  & \multicolumn{1}{c}{\textbf{\boldmath$\sigma_{\text{sys}}$}}  & & \multicolumn{1}{c}{\textbf{Parameter}} & \multicolumn{1}{c}{\textbf{Value}}  & \multicolumn{1}{c}{\textbf{\boldmath$\sigma_{\text{stat}}$}}  & \multicolumn{1}{c}{\textbf{\boldmath$\sigma_{\text{sys}}$}} \\ \midrule
$J_0$ (\SI{}{\kilo\meter^{-2} \steradian^{-1} \year^{-1} \eV^{-1}}) & 1.269 & 0.003 & 0.40  & & $\gamma_1$ & 3.26 & 0.01 & 0.10 \\
$E_1$ (\SI{}{\EeV})  & 5.1 & 0.1 & 1.1 & & $\gamma_2$ & 2.51 & 0.03 & 0.05  \\
$E_2$ (\SI{}{\EeV}) & 13 & 1 & 2 & & $\gamma_3$ & 2.99 & 0.03 & 0.10 \\
$E_3$ (\SI{}{\EeV}) & 48 & 2 & 5 & & $\gamma_4$ & 5.4 & 0.2 & 0.1  \\
\bottomrule
\end{tabular}
\label{tab:spectral_parameters}
\end{table}

\section{Phase I spectrum}
\label{sec:combined_spectrum}

Given that the spectra in the different declination bands agree with the flux fitted in the common declination band within statistical uncertainties, we combine the vertical and inclined into a single spectrum.
While in section~\ref{sec:declination} we restricted the vertical and inclined spectra to their common declination band $[-\SI{84.8}{\degree}, +\SI{24.8}{\degree}]$, we now combine them at all observed declinations, reaching a total exposure of \SI{104900}{\square\kilo\meter\steradian\year} corresponding to the sum of the exposure of the vertical and inclined datasets.
The combined spectrum is calculated using Eq.~\ref{eq:combined_spectrum} after shifting the inclined spectrum with the parameters calculated in section~\ref{sec:declination}. 
We show the combined spectrum and the corresponding fit in Fig.~\ref{fig:combined_spectrum}, and the spectrum data are provided in~\cite{PierreAuger:2025hnw}. 
The shaded band represents the systematic uncertainty of the flux, which is dominated by the systematic uncertainty of the energy scale.
The flux exhibits the firmly established features of the ankle and suppression, as well as the instep, which we unveiled with a significance of 3.9$\sigma$ in 2020~\cite{PierreAuger:2020qqz}.
Table~\ref{tab:spectral_parameters} contains the fitted spectral parameters and their statistical and systematic uncertainties.

\noindent
\begin{minipage}[t]{0.48\textwidth}
  \begin{figure}[H]
    \centering
    \includegraphics[width=\linewidth]{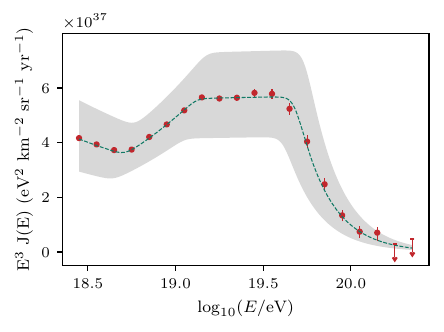}
    \caption{Combined spectrum using events arriving with zenith angle up to \SI{80}{\degree} observed at declinations from the south celestial pole up to +\SI{44.8}{\degree}. The systematic flux uncertainty is shown as a shaded band.}
\label{fig:combined_spectrum} \end{figure}
\end{minipage}
\hfill
\begin{minipage}[t]{0.48\textwidth}
  \begin{figure}[H]
    \centering
    \includegraphics[width=\linewidth]{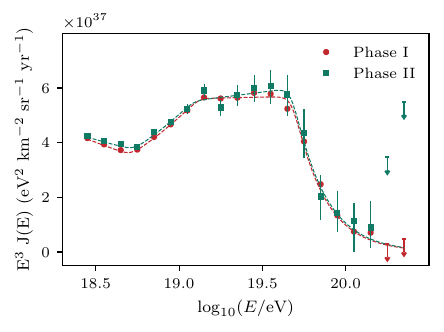}
    \caption{Preliminary spectrum using events arriving with zenith angle up to \SI{60}{\degree} observed during the second phase of Auger. The spectrum of Phase I is shown for comparison.}
    \label{fig:phase2}
  \end{figure}
\end{minipage}
\vspace{2em}

To assess the significance of the instep with the larger dataset currently available, we sampled energies from a reference model, which contained a slow suppression instead of the instep.
We fitted the sampled data with the reference model and alternative one containing the instep and built a test statistic as the ratio of their respective likelihoods.
The estimated significance is based on the number of times the sampled test statistic is greater than the one calculated by fitting the observed data.  
The test statistic was larger than the observed value ($t_{\text{obs}}~\sim35$) in only two out of $10^8$ simulations, corresponding to a significance of 5.5$\sigma$.

\section{Preliminary Phase II spectrum}

In this section, we present a preliminary spectrum built from vertical events observed during the second Phase of the Auger Observatory, using data recorded from April 1, 2023, to March 1, 2025.
To provide this first view of the Phase II spectrum, we reconstructed the events using the Phase I setup. 
For example, we applied the same lateral distribution function, attenuation with zenith angle, energy calibration, and detector response model used to unfold the spectrum.
We also use the same method as in Phase I to select spectrum events, excluding periods of unstable data-taking, and to calculate the exposure.
The exposure during the data-taking considered was \SI{9200 \pm 300}{\square\kilo\meter\steradian\year}, about $10\%$ of the Phase I spectrum.
Reusing the Phase I analysis was possible because the upgrade of the Observatory was designed so that water-Cherenkov detectors were backwards compatible.

We show in Fig.~\ref{fig:phase2} the spectra of Phase I and II.
We evaluated the consistency of Phase I and Phase II spectra with a Fisher's test that combines chi-square tests for bins with many events with exact binomial tests for bins with few events~\cite{Ravignani:2025gqu}. 
We assumed a systematic uncertainty of $1\%$ uncorrelated between the two spectra, a conservative estimate given the $3\%$ exposure uncertainty.
The observed Fisher's test statistic was $t=38.0$ for 36 degrees of freedom, corresponding to a p-value $p=0.30$ estimated numerically with Monte Carlo simulations. 

We fitted the Phase II spectrum with the Phase I model, and obtained spectral parameters that are in agreement with the values in Table~\ref{tab:spectral_parameters} within the statistical uncertainties. 
We will customize the event reconstruction to accommodate the improvements of the water-Cherenkov detectors in Phase II, but this early analysis already shows that minor changes to the existing setup will be required.
In the future, we will use Phase I and II events to build a single spectrum with the combined exposure achieved during the complete lifetime of the Pierre Auger Observatory.

\section{Conclusions}

We presented an updated spectrum measured with the \SI{1500}{\meter} Surface Detector Array of the Pierre Auger Observatory using events arriving with zenith angles up to $\SI{80}{\degree}$. 
The data set comprises 19 years of Phase I data taken from 2004 to 2023 with an exposure of \SI{104900 \pm 3000}{\square\kilo\meter\steradian\year}, enabling the discovery-level observation of the spectrum instep at \SI{13 \pm 1 \pm 2}{\EeV} with the spectral index increasing from $2.51 \pm 0.03^\text{stat} \pm 0.05^\text{sys}$ to $2.99 \pm 0.03^\text{stat} \pm 0.10^\text{sys}$.
We did not find any statistically significant dependence of the flux with declination from the south celestial pole up to a $+\SI{44.8}{\degree}$, bar a small trend consistent with the well-established dipolar anisotropy in arrival directions.
Finally, we presented for the first time a spectrum obtained with events measured during Phase II of the Auger Observatory, showing its consistency with the Phase I spectrum.

\begingroup
\small
\setstretch{0.5}  
\bibliographystyle{apsrev4-2}
\bibliography{biblio}
\endgroup

\newpage
\section*{The Pierre Auger Collaboration}

{\footnotesize\setlength{\baselineskip}{10pt}
\noindent
\begin{wrapfigure}[11]{l}{0.12\linewidth}
\vspace{-4pt}
\includegraphics[width=0.98\linewidth]{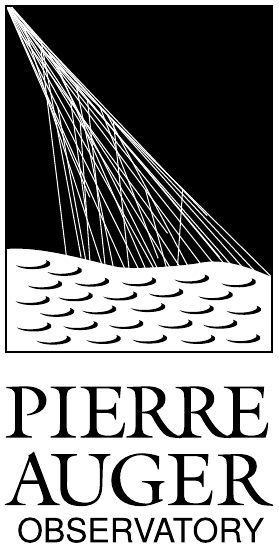}
\end{wrapfigure}
\begin{sloppypar}\noindent

\noindent A.~Abdul Halim$^{13}$,
P.~Abreu$^{70}$,
M.~Aglietta$^{53,51}$,
I.~Allekotte$^{1}$,
K.~Almeida Cheminant$^{78,77}$,
A.~Almela$^{7,12}$,
R.~Aloisio$^{44,45}$,
J.~Alvarez-Mu\~niz$^{76}$,
A.~Ambrosone$^{44}$,
J.~Ammerman Yebra$^{76}$
% ...

\end{sloppypar}
\begin{center}
\end{center}

\vspace{1ex}
\begin{description}[labelsep=0.2em,align=right,labelwidth=0.7em,labelindent=0em,leftmargin=2em,noitemsep,before={\renewcommand\makelabel[1]{##1 }}]
\item[$^{1}$] Centro At\'omico Bariloche and Instituto Balseiro (CNEA-UNCuyo-CONICET), San Carlos de Bariloche, Argentina
\item[$^{2}$] Departamento de F\'\i{}sica and Departamento de Ciencias de la Atm\'osfera y los Oc\'eanos, FCEyN, Universidad de Buenos Aires and CONICET, Buenos Aires, Argentina
\item[$^{3}$] IFLP, Universidad Nacional de La Plata and CONICET, La Plata, Argentina
\item[$^{4}$] Instituto de Astronom\'\i{}a y F\'\i{}sica del Espacio (IAFE, CONICET-UBA), Buenos Aires, Argentina
\item[$^{5}$] Instituto de F\'\i{}sica de Rosario (IFIR) -- CONICET/U.N.R.\ and Facultad de Ciencias Bioqu\'\i{}micas y Farmac\'euticas U.N.R., Rosario, Argentina
\item[$^{6}$] Instituto de Tecnolog\'\i{}as en Detecci\'on y Astropart\'\i{}culas (CNEA, CONICET, UNSAM), and Universidad Tecnol\'ogica Nacional -- Facultad Regional Mendoza (CONICET/CNEA), Mendoza, Argentina
\item[$^{7}$] Instituto de Tecnolog\'\i{}as en Detecci\'on y Astropart\'\i{}culas (CNEA, CONICET, UNSAM), Buenos Aires, Argentina
\item[$^{8}$] International Center of Advanced Studies and Instituto de Ciencias F\'\i{}sicas, ECyT-UNSAM and CONICET, Campus Miguelete -- San Mart\'\i{}n, Buenos Aires, Argentina
\item[$^{9}$] Laboratorio Atm\'osfera -- Departamento de Investigaciones en L\'aseres y sus Aplicaciones -- UNIDEF (CITEDEF-CONICET), Argentina
\item[$^{10}$] Observatorio Pierre Auger, Malarg\"ue, Argentina
\item[$^{11}$] Observatorio Pierre Auger and Comisi\'on Nacional de Energ\'\i{}a At\'omica, Malarg\"ue, Argentina
\item[$^{12}$] Universidad Tecnol\'ogica Nacional -- Facultad Regional Buenos Aires, Buenos Aires, Argentina
\item[$^{13}$] University of Adelaide, Adelaide, S.A., Australia
\item[$^{14}$] Universit\'e Libre de Bruxelles (ULB), Brussels, Belgium
\item[$^{15}$] Vrije Universiteit Brussels, Brussels, Belgium
\item[$^{16}$] Centro Brasileiro de Pesquisas Fisicas, Rio de Janeiro, RJ, Brazil
\item[$^{17}$] Centro Federal de Educa\c{c}\~ao Tecnol\'ogica Celso Suckow da Fonseca, Petropolis, Brazil
\item[$^{18}$] Instituto Federal de Educa\c{c}\~ao, Ci\^encia e Tecnologia do Rio de Janeiro (IFRJ), Brazil
\item[$^{19}$] Universidade de S\~ao Paulo, Escola de Engenharia de Lorena, Lorena, SP, Brazil
\item[$^{20}$] Universidade de S\~ao Paulo, Instituto de F\'\i{}sica de S\~ao Carlos, S\~ao Carlos, SP, Brazil
\item[$^{21}$] Universidade de S\~ao Paulo, Instituto de F\'\i{}sica, S\~ao Paulo, SP, Brazil
\item[$^{22}$] Universidade Estadual de Campinas (UNICAMP), IFGW, Campinas, SP, Brazil
\item[$^{23}$] Universidade Estadual de Feira de Santana, Feira de Santana, Brazil
\item[$^{24}$] Universidade Federal de Campina Grande, Centro de Ciencias e Tecnologia, Campina Grande, Brazil
\item[$^{25}$] Universidade Federal do ABC, Santo Andr\'e, SP, Brazil
\item[$^{26}$] Universidade Federal do Paran\'a, Setor Palotina, Palotina, Brazil
\item[$^{27}$] Universidade Federal do Rio de Janeiro, Instituto de F\'\i{}sica, Rio de Janeiro, RJ, Brazil
\item[$^{28}$] Universidad de Medell\'\i{}n, Medell\'\i{}n, Colombia
\item[$^{29}$] Universidad Industrial de Santander, Bucaramanga, Colombia
\item[$^{30}$] Charles University, Faculty of Mathematics and Physics, Institute of Particle and Nuclear Physics, Prague, Czech Republic
\item[$^{31}$] Institute of Physics of the Czech Academy of Sciences, Prague, Czech Republic
\item[$^{32}$] Palacky University, Olomouc, Czech Republic
\item[$^{33}$] CNRS/IN2P3, IJCLab, Universit\'e Paris-Saclay, Orsay, France
\item[$^{34}$] Laboratoire de Physique Nucl\'eaire et de Hautes Energies (LPNHE), Sorbonne Universit\'e, Universit\'e de Paris, CNRS-IN2P3, Paris, France
\item[$^{35}$] Univ.\ Grenoble Alpes, CNRS, Grenoble Institute of Engineering Univ.\ Grenoble Alpes, LPSC-IN2P3, 38000 Grenoble, France
\item[$^{36}$] Universit\'e Paris-Saclay, CNRS/IN2P3, IJCLab, Orsay, France
\item[$^{37}$] Bergische Universit\"at Wuppertal, Department of Physics, Wuppertal, Germany
\item[$^{38}$] Karlsruhe Institute of Technology (KIT), Institute for Experimental Particle Physics, Karlsruhe, Germany
\item[$^{39}$] Karlsruhe Institute of Technology (KIT), Institut f\"ur Prozessdatenverarbeitung und Elektronik, Karlsruhe, Germany
\item[$^{40}$] Karlsruhe Institute of Technology (KIT), Institute for Astroparticle Physics, Karlsruhe, Germany
\item[$^{41}$] RWTH Aachen University, III.\ Physikalisches Institut A, Aachen, Germany
\item[$^{42}$] Universit\"at Hamburg, II.\ Institut f\"ur Theoretische Physik, Hamburg, Germany
\item[$^{43}$] Universit\"at Siegen, Department Physik -- Experimentelle Teilchenphysik, Siegen, Germany
\item[$^{44}$] Gran Sasso Science Institute, L'Aquila, Italy
\item[$^{45}$] INFN Laboratori Nazionali del Gran Sasso, Assergi (L'Aquila), Italy
\item[$^{46}$] INFN, Sezione di Catania, Catania, Italy
\item[$^{47}$] INFN, Sezione di Lecce, Lecce, Italy
\item[$^{48}$] INFN, Sezione di Milano, Milano, Italy
\item[$^{49}$] INFN, Sezione di Napoli, Napoli, Italy
\item[$^{50}$] INFN, Sezione di Roma ``Tor Vergata'', Roma, Italy
\item[$^{51}$] INFN, Sezione di Torino, Torino, Italy
\item[$^{52}$] Istituto di Astrofisica Spaziale e Fisica Cosmica di Palermo (INAF), Palermo, Italy
\item[$^{53}$] Osservatorio Astrofisico di Torino (INAF), Torino, Italy
\item[$^{54}$] Politecnico di Milano, Dipartimento di Scienze e Tecnologie Aerospaziali , Milano, Italy
\item[$^{55}$] Universit\`a del Salento, Dipartimento di Matematica e Fisica ``E.\ De Giorgi'', Lecce, Italy
\item[$^{56}$] Universit\`a dell'Aquila, Dipartimento di Scienze Fisiche e Chimiche, L'Aquila, Italy
\item[$^{57}$] Universit\`a di Catania, Dipartimento di Fisica e Astronomia ``Ettore Majorana``, Catania, Italy
\item[$^{58}$] Universit\`a di Milano, Dipartimento di Fisica, Milano, Italy
\item[$^{59}$] Universit\`a di Napoli ``Federico II'', Dipartimento di Fisica ``Ettore Pancini'', Napoli, Italy
\item[$^{60}$] Universit\`a di Palermo, Dipartimento di Fisica e Chimica ''E.\ Segr\`e'', Palermo, Italy
\item[$^{61}$] Universit\`a di Roma ``Tor Vergata'', Dipartimento di Fisica, Roma, Italy
\item[$^{62}$] Universit\`a Torino, Dipartimento di Fisica, Torino, Italy
\item[$^{63}$] Benem\'erita Universidad Aut\'onoma de Puebla, Puebla, M\'exico
\item[$^{64}$] Unidad Profesional Interdisciplinaria en Ingenier\'\i{}a y Tecnolog\'\i{}as Avanzadas del Instituto Polit\'ecnico Nacional (UPIITA-IPN), M\'exico, D.F., M\'exico
\item[$^{65}$] Universidad Aut\'onoma de Chiapas, Tuxtla Guti\'errez, Chiapas, M\'exico
\item[$^{66}$] Universidad Michoacana de San Nicol\'as de Hidalgo, Morelia, Michoac\'an, M\'exico
\item[$^{67}$] Universidad Nacional Aut\'onoma de M\'exico, M\'exico, D.F., M\'exico
\item[$^{68}$] Institute of Nuclear Physics PAN, Krakow, Poland
\item[$^{69}$] University of \L{}\'od\'z, Faculty of High-Energy Astrophysics,\L{}\'od\'z, Poland
\item[$^{70}$] Laborat\'orio de Instrumenta\c{c}\~ao e F\'\i{}sica Experimental de Part\'\i{}culas -- LIP and Instituto Superior T\'ecnico -- IST, Universidade de Lisboa -- UL, Lisboa, Portugal
\item[$^{71}$] ``Horia Hulubei'' National Institute for Physics and Nuclear Engineering, Bucharest-Magurele, Romania
\item[$^{72}$] Institute of Space Science, Bucharest-Magurele, Romania
\item[$^{73}$] Center for Astrophysics and Cosmology (CAC), University of Nova Gorica, Nova Gorica, Slovenia
\item[$^{74}$] Experimental Particle Physics Department, J.\ Stefan Institute, Ljubljana, Slovenia
\item[$^{75}$] Universidad de Granada and C.A.F.P.E., Granada, Spain
\item[$^{76}$] Instituto Galego de F\'\i{}sica de Altas Enerx\'\i{}as (IGFAE), Universidade de Santiago de Compostela, Santiago de Compostela, Spain
\item[$^{77}$] IMAPP, Radboud University Nijmegen, Nijmegen, The Netherlands
\item[$^{78}$] Nationaal Instituut voor Kernfysica en Hoge Energie Fysica (NIKHEF), Science Park, Amsterdam, The Netherlands
\item[$^{79}$] Stichting Astronomisch Onderzoek in Nederland (ASTRON), Dwingeloo, The Netherlands
\item[$^{80}$] Universiteit van Amsterdam, Faculty of Science, Amsterdam, The Netherlands
\item[$^{81}$] Case Western Reserve University, Cleveland, OH, USA
\item[$^{82}$] Colorado School of Mines, Golden, CO, USA
\item[$^{83}$] Department of Physics and Astronomy, Lehman College, City University of New York, Bronx, NY, USA
\item[$^{84}$] Michigan Technological University, Houghton, MI, USA
\item[$^{85}$] New York University, New York, NY, USA
\item[$^{86}$] University of Chicago, Enrico Fermi Institute, Chicago, IL, USA
\item[$^{87}$] University of Delaware, Department of Physics and Astronomy, Bartol Research Institute, Newark, DE, USA
\item[] -----
\item[$^{a}$] Max-Planck-Institut f\"ur Radioastronomie, Bonn, Germany
\item[$^{b}$] also at Kapteyn Institute, University of Groningen, Groningen, The Netherlands
\item[$^{c}$] School of Physics and Astronomy, University of Leeds, Leeds, United Kingdom
\item[$^{d}$] Fermi National Accelerator Laboratory, Fermilab, Batavia, IL, USA
\item[$^{e}$] Pennsylvania State University, University Park, PA, USA
\item[$^{f}$] Colorado State University, Fort Collins, CO, USA
\item[$^{g}$] Louisiana State University, Baton Rouge, LA, USA
\item[$^{h}$] now at Graduate School of Science, Osaka Metropolitan University, Osaka, Japan
\item[$^{i}$] Institut universitaire de France (IUF), France
\item[$^{j}$] now at Technische Universit\"at Dortmund and Ruhr-Universit\"at Bochum, Dortmund and Bochum, Germany
\end{description}

\section*{Acknowledgments}

\begin{sloppypar}
The successful installation, commissioning, and operation of the Pierre
Auger Observatory would not have been possible without the strong
commitment and effort from the technical and administrative staff in
Malarg\"ue. We are very grateful to the following agencies and
organizations for financial support:
\end{sloppypar}

\begin{sloppypar}
Argentina -- Comisi\'on Nacional de Energ\'\i{}a At\'omica; Agencia Nacional de
Promoci\'on Cient\'\i{}fica y Tecnol\'ogica (ANPCyT); Consejo Nacional de
Investigaciones Cient\'\i{}ficas y T\'ecnicas (CONICET); Gobierno de la
Provincia de Mendoza; Municipalidad de Malarg\"ue; NDM Holdings and Valle
Las Le\~nas; in gratitude for their continuing cooperation over land
access; Australia -- the Australian Research Council; Belgium -- Fonds
de la Recherche Scientifique (FNRS); Research Foundation Flanders (FWO),
Marie Curie Action of the European Union Grant No.~101107047; Brazil --
Conselho Nacional de Desenvolvimento Cient\'\i{}fico e Tecnol\'ogico (CNPq);
Financiadora de Estudos e Projetos (FINEP); Funda\c{c}\~ao de Amparo \`a
Pesquisa do Estado de Rio de Janeiro (FAPERJ); S\~ao Paulo Research
Foundation (FAPESP) Grants No.~2019/10151-2, No.~2010/07359-6 and
No.~1999/05404-3; Minist\'erio da Ci\^encia, Tecnologia, Inova\c{c}\~oes e
Comunica\c{c}\~oes (MCTIC); Czech Republic -- GACR 24-13049S, CAS LQ100102401,
MEYS LM2023032, CZ.02.1.01/0.0/0.0/16{\textunderscore}013/0001402,
CZ.02.1.01/0.0/0.0/18{\textunderscore}046/0016010 and
CZ.02.1.01/0.0/0.0/17{\textunderscore}049/0008422 and CZ.02.01.01/00/22{\textunderscore}008/0004632;
France -- Centre de Calcul IN2P3/CNRS; Centre National de la Recherche
Scientifique (CNRS); Conseil R\'egional Ile-de-France; D\'epartement
Physique Nucl\'eaire et Corpusculaire (PNC-IN2P3/CNRS); D\'epartement
Sciences de l'Univers (SDU-INSU/CNRS); Institut Lagrange de Paris (ILP)
Grant No.~LABEX ANR-10-LABX-63 within the Investissements d'Avenir
Programme Grant No.~ANR-11-IDEX-0004-02; Germany -- Bundesministerium
f\"ur Bildung und Forschung (BMBF); Deutsche Forschungsgemeinschaft (DFG);
Finanzministerium Baden-W\"urttemberg; Helmholtz Alliance for
Astroparticle Physics (HAP); Helmholtz-Gemeinschaft Deutscher
Forschungszentren (HGF); Ministerium f\"ur Kultur und Wissenschaft des
Landes Nordrhein-Westfalen; Ministerium f\"ur Wissenschaft, Forschung und
Kunst des Landes Baden-W\"urttemberg; Italy -- Istituto Nazionale di
Fisica Nucleare (INFN); Istituto Nazionale di Astrofisica (INAF);
Ministero dell'Universit\`a e della Ricerca (MUR); CETEMPS Center of
Excellence; Ministero degli Affari Esteri (MAE), ICSC Centro Nazionale
di Ricerca in High Performance Computing, Big Data and Quantum
Computing, funded by European Union NextGenerationEU, reference code
CN{\textunderscore}00000013; M\'exico -- Consejo Nacional de Ciencia y Tecnolog\'\i{}a
(CONACYT) No.~167733; Universidad Nacional Aut\'onoma de M\'exico (UNAM);
PAPIIT DGAPA-UNAM; The Netherlands -- Ministry of Education, Culture and
Science; Netherlands Organisation for Scientific Research (NWO); Dutch
national e-infrastructure with the support of SURF Cooperative; Poland
-- Ministry of Education and Science, grants No.~DIR/WK/2018/11 and
2022/WK/12; National Science Centre, grants No.~2016/22/M/ST9/00198,
2016/23/B/ST9/01635, 2020/39/B/ST9/01398, and 2022/45/B/ST9/02163;
Portugal -- Portuguese national funds and FEDER funds within Programa
Operacional Factores de Competitividade through Funda\c{c}\~ao para a Ci\^encia
e a Tecnologia (COMPETE); Romania -- Ministry of Research, Innovation
and Digitization, CNCS-UEFISCDI, contract no.~30N/2023 under Romanian
National Core Program LAPLAS VII, grant no.~PN 23 21 01 02 and project
number PN-III-P1-1.1-TE-2021-0924/TE57/2022, within PNCDI III; Slovenia
-- Slovenian Research Agency, grants P1-0031, P1-0385, I0-0033, N1-0111;
Spain -- Ministerio de Ciencia e Innovaci\'on/Agencia Estatal de
Investigaci\'on (PID2019-105544GB-I00, PID2022-140510NB-I00 and
RYC2019-027017-I), Xunta de Galicia (CIGUS Network of Research Centers,
Consolidaci\'on 2021 GRC GI-2033, ED431C-2021/22 and ED431F-2022/15),
Junta de Andaluc\'\i{}a (SOMM17/6104/UGR and P18-FR-4314), and the European
Union (Marie Sklodowska-Curie 101065027 and ERDF); USA -- Department of
Energy, Contracts No.~DE-AC02-07CH11359, No.~DE-FR02-04ER41300,
No.~DE-FG02-99ER41107 and No.~DE-SC0011689; National Science Foundation,
Grant No.~0450696, and NSF-2013199; The Grainger Foundation; Marie
Curie-IRSES/EPLANET; European Particle Physics Latin American Network;
and UNESCO.
\end{sloppypar}

}

\end{document}